\pdfoutput=1
\documentclass[11pt]{article}
\usepackage[margin=1.1in]{geometry}
\usepackage[T1]{fontenc}
\usepackage{lmodern}
\usepackage{microtype}
\usepackage{booktabs}
\usepackage{listings}
\usepackage{xcolor}
\usepackage{enumitem}
\usepackage{amsmath}
\usepackage{tikz}
\usetikzlibrary{positioning,arrows.meta}
\usepackage{hyperref}

\hypersetup{colorlinks=true, linkcolor=blue!50!black, citecolor=blue!50!black, urlcolor=blue!50!black}

\title{NEBULA: A Language-Independent Specification for\\Opaque Rotating Refresh Tokens}
\author{Matteo Teodori\\\small Independent researcher\\\small
  \href{mailto:hello@nebulatoken.dev}{hello@nebulatoken.dev}}
\date{4 August 2026}

\begin{document}
\maketitle

\begin{abstract}
Refresh tokens are among the most sensitive credentials in modern
authentication systems: long-lived, bearer-style, and sufficient to mint
access tokens for days or weeks. RFC~9700, the current Best Current Practice
for OAuth~2.0 security, mandates that refresh tokens issued to public clients
be rotated on every use with replay (reuse) detection, or be
sender-constrained. However, the BCP
specifies \emph{policy}, not \emph{mechanism}: it prescribes no wire format,
no storage schema, no ordering of verification steps, no concurrency
contract, and no semantics for edge cases such as lost-response retries or
key rotation. As a result, production implementations---both in managed
identity providers and in first-party authentication code---may diverge in
precisely the corner cases that determine security outcomes.

We present NEBULA, a precise, language-independent specification of the
RFC~9700 refresh-token model, together with ten conformant reference
implementations (TypeScript, Python, Go, Rust, Java, PHP, C\#, Ruby, Elixir,
Dart). NEBULA tokens are opaque---a 128-bit public selector and a 256-bit
secret verifier, both CSPRNG output, carrying no claims and no
signature---so token validity is a property of server-side
state rather than of cryptographic verification. Its conformance methodology
publishes the behavioural suite \emph{as data} rather than as prose: 38
scenarios in one machine-readable file that every implementation executes
through a thin per-language runner, so that drift by transcription is
structurally excluded rather than merely unintended, and divergence at the
observation points the vectors define is a failing test rather than a
discovery made in production. We
describe the specification---including a compare-and-set rotation contract
that closes a reproducible bypass of reuse detection under concurrent
refresh---analyse its security properties including its post-quantum
posture, and report on cross-language conformance as a methodology for
multi-implementation security specifications. The specification,
implementations, and conformance artefacts are open source under the Apache
License 2.0.
\end{abstract}

\noindent\textbf{ACM classification.}\; K.6.5 [\emph{Management of Computing
and Information Systems}]: Security and Protection (primary);
E.3 [\emph{Data Encryption}] (secondary).

\section{Introduction}

Token-based authentication separates credentials into two roles: short-lived
\emph{access tokens} presented to resource servers on every request, and
long-lived \emph{refresh tokens} exchanged periodically for fresh access
tokens. This separation, standardized by OAuth~2.0~\cite{rfc6749}, is now
ubiquitous well beyond delegated authorization: most first-party web and
mobile applications implement the same pattern.

The security asymmetry between the two roles is stark. An access token
expires in minutes; a refresh token is typically valid for days to weeks and
is, in the common deployment, a pure bearer credential: whoever presents it
obtains authenticated access. Exfiltration of a refresh token---through
cross-site scripting, device malware, leaked logs, or backup
exposure---therefore grants an attacker persistent, silent access for the
remaining token lifetime.

RFC~9700~\cite{rfc9700} addresses this threat by requiring that refresh
tokens for public clients either be sender-constrained (e.g., via
DPoP~\cite{rfc9449} or mutual TLS~\cite{rfc8705}) or be \emph{rotated} on
every use with \emph{replay detection}: if a previously-used refresh token is
presented again, two parties demonstrably hold the same credential, and the
authorization server revokes the active refresh token. NEBULA strengthens that
last step to revocation of the entire rotation lineage (\S\ref{sec:reuse}); the
BCP does not require it, and reserves its set-wide vocabulary for a different
case, directing that an authorization server ``SHOULD revoke all tokens'' issued
on a replayed authorization code.

\paragraph{The specification gap.}
RFC~9700 is deliberately a policy document. It does not define: a token
format; which party's keys verify a presented token after key rotation; the
order of verification checks (and hence which error is observable when
multiple conditions fail); whether and how a legitimate client that lost a
rotation response may retry; what state must be persisted, and for how long,
for replay detection to function; what happens when two refreshes of one
token execute concurrently; or how session lifetimes interact with rotation.
Each implementer answers these questions independently. Managed identity
providers implement rotation and reuse detection internally, under semantics
they are not obliged to publish or to share with one another; first-party
implementations answer from first principles.

We make no empirical claim here about how often that goes wrong, having no
survey to support one. We observe instead that a document specifying
rotation-with-reuse-detection at the level of policy does not exclude an
implementation with any of the following properties, each of which silently
destroys the guarantee the policy is meant to deliver, and each of which the
present specification forbids by an identified requirement:

\begin{itemize}[leftmargin=1.4em, itemsep=2pt, topsep=3pt]
  \item a replay window extendable without bound by repeated retries;
  \item a store that expires rotated records under an ordinary TTL,
  converting every replay from a detected reuse into a benign
  ``not found'' (\S\ref{sec:state});
  \item a rotation path without a concurrency contract, under which two
  simultaneous refreshes of one token fork a family into two independently
  valid lineages and no later presentation is ever a replay
  (\S\ref{sec:cas});
  \item revocation authorised by the public lookup key alone, turning a
  database read or a log line into the ability to terminate an arbitrary
  session (\S\ref{sec:revoke}).
\end{itemize}

\paragraph{Contributions.} This paper makes three contributions:
\begin{enumerate}[leftmargin=1.4em]
  \item \textbf{A precise specification} (Section~\ref{sec:spec}) of the
  rotation-with-reuse-detection model: an opaque token format with a
  selector/verifier split, a six-method storage contract whose two
  transitions out of \textit{active} that can race are compare-and-set, a ten-step
  verification algorithm with fixed and observable ordering, bounded
  non-extendable retry semantics, normative record retention, authenticated
  revocation by token, and zero-downtime key rotation via key identifiers.
  \item \textbf{A security analysis} (Section~\ref{sec:security}) arguing
  that the design converts refresh-token theft from a silent, long-lived
  compromise into a bounded and---outside one explicitly documented
  window---detectable event, renders a full database disclosure
  non-actionable both for authentication and for revocation, closes a
  reproducible bypass of reuse detection under concurrent rotation, and,
  containing no public-key cryptography, requires no post-quantum migration
  of the token layer.
  \item \textbf{A conformance methodology} (Section~\ref{sec:conformance})
  for multi-language security specifications, in which both conformance
  artefacts are published as data: 46 deterministic vectors fixing the keyed
  primitives and the parser byte-for-byte, and a behavioural suite of 38
  scenarios---sequences of engine operations against an injected clock, with
  their expected outcomes and resulting record states---in a single
  machine-readable file that all ten implementations execute through a thin
  per-language runner. No implementation restates the suite in its own test
  framework, so the ten cannot drift apart by transcription. The published
  case counts are themselves normative: a runner that silently executes zero
  cases fails rather than passes.
\end{enumerate}

We explicitly claim no cryptographic novelty. NEBULA composes
HMAC-SHA-256~\cite{rfc2104,fips198}, CSPRNG output, and constant-time
comparison in a deliberately conventional manner; its contribution is the
elimination of underspecification, which we argue is the dominant source of
vulnerabilities in this layer.

\section{Background and Related Work}

\paragraph{Opaque session identifiers.} Server-side sessions keyed by
unguessable identifiers predate token-based architectures and share NEBULA's
core property: validity as server-side state. NEBULA can be viewed as a
systematization of this classical model, extended with rotation lineage,
replay detection, split lookup/proof material, and peppered at-rest hashing.

\paragraph{Self-contained tokens.} JWT~\cite{rfc7519} and its profile for
OAuth access tokens~\cite{rfc9068} embed claims and a signature in the
token, enabling stateless verification by third parties.
PASETO~\cite{paseto} and Biscuit~\cite{biscuit} refine the same idea with
restricted algorithm choices and offline attenuation respectively.
Self-contained designs are appropriate where many verifiers must validate
tokens at high frequency; they are structurally unable to provide immediate
revocation, which is the defining requirement for refresh tokens. NEBULA is
complementary to these designs, not competitive with them.

\paragraph{Sender-constrained tokens.} DPoP~\cite{rfc9449} and certificate-bound
tokens~\cite{rfc8705} bind tokens to client-held keys, defeating bearer-token
replay entirely at the cost of client-side key management. RFC~9700 treats
sender-constraining and rotation-with-reuse-detection as alternative
mitigations. NEBULA implements the latter and provides an application-level
device-binding hook; cryptographic sender-constraining can be layered above
it (\S\ref{sec:limitations}).

\paragraph{Split lookup/proof credentials.} Separating a stable lookup key
from a rotating proof value is long-standing practice for persistent-login
(``remember-me'') cookies: Jaspan pairs a reused \emph{series} identifier with
a single-use token so that a valid series presented with a stale token betrays
a theft~\cite{jaspan2006} --- an early statement of the reuse-detection
argument this paper develops in \S\ref{sec:reuse}, although the tokens there
are stored in the clear. The refinement NEBULA adopts---treating the lookup key
as public and storing only a keyed hash of the proof value, so that what the
lookup's timing can reveal is by construction the half that is not
secret---is the split-token construction of Paragon Initiative
Enterprises~\cite{paragonie-split}. NEBULA adopts it normatively.

\section{Threat Model}
\label{sec:threat}

We consider an adversary who may: (T1) obtain a copy of any token in transit
or at rest on the client; (T2) read a complete copy of the server-side token
database, excluding process environment and key-management systems; (T3)
submit arbitrary strings to the refresh endpoint; (T4) measure response
timing. We additionally consider (T5) an adversary or an ordinary client
whose requests \emph{interleave}, so that two refreshes of one token execute
concurrently; this is not an exotic capability but the default behaviour of
retrying mobile clients and multi-tab browsers. We assume the server
process, its environment (peppers), and the CSPRNG are uncompromised;
transport security (TLS) is assumed for confidentiality in transit. Denial
of service against the backing store and full client compromise after login
are out of scope, as for any token scheme.

\section{The NEBULA Specification}
\label{sec:spec}

This section summarizes the normative specification; the authoritative text
and the shared conformance artefacts are published in the project
repository~\cite{nebularepo}. Requirement keywords are used as defined in
RFC~2119~\cite{rfc2119} and RFC~8174~\cite{rfc8174}, and every requirement
in the normative text is individually identified as [N-\emph{n}], so that
tests, threat-model entries and third-party conformance reports can cite an
exact obligation rather than a section. There are 53 such requirements in
version~1 of the specification; we cite the load-bearing ones below by
identifier.

\subsection{Token Format}

A NEBULA token is the ASCII string
\begin{lstlisting}
nbl.{kid}.{selector}.{verifier}
\end{lstlisting}
where \texttt{kid} identifies the server-side HMAC key (\emph{pepper}) used
for at-rest hashing; \texttt{selector} is the base64url encoding
(RFC~4648~\S5, unpadded~\cite{rfc4648}) of 16 CSPRNG bytes and serves as the
public database key; and \texttt{verifier} is the base64url encoding of 32
CSPRNG bytes and constitutes the secret --- comfortably above the 64 bits of
CSPRNG entropy OWASP requires of a session
identifier~\cite{owasp-session}. Tokens exceeding 512 bytes,
containing other than four non-empty dot-separated parts, or whose verifier
fails to decode to exactly 32 bytes under \emph{canonical} encoding, are
rejected before any database interaction [N-6]. The canonical-encoding
requirement guarantees one wire form per credential even under permissive
base64 decoders: a 32-byte value has four distinct 43-character base64url
encodings, because the final character carries four significant bits and two
unused ones, and without the requirement a permissive decoder in one
implementation and a strict decoder in another disagree on whether a
presented string is a token at all [N-7].

The token carries no claims, no expiry, and no signature: it is
distinguishable from random only by its framing.

\subsection{Server-Side State and Storage Contract}
\label{sec:state}

One record per issued token stores the selector (primary key), a peppered
hash of the verifier, the kid, a 128-bit random \emph{family} identifier
shared by all rotations of one login, a generation counter, the user
identifier, an optional peppered device-binding hash, the creation time, a
\emph{family expiry} fixed at login and never extended, a sliding \emph{idle
expiry}, a status in $\{\textit{active}, \textit{rotated},
\textit{revoked}\}$, and---once rotated---the rotation time and the
successor selector [N-10]. The two hashes are
\begin{equation*}
  h_v = \mathrm{HMAC}(\mathit{pepper}_{kid},\, \mathit{verifier}),
  \qquad
  h_d = \mathrm{HMAC}(\mathit{pepper}_{kid},\, \texttt{"device:"} \,\|\, d),
\end{equation*}
both HMAC-SHA-256 rendered as lowercase hex, where $\mathit{pepper}_{kid}$
is the pepper named by the \emph{record's} own kid and $d$ is the device
identifier [N-11, N-13]. Raw verifiers and raw device identifiers are never
persisted, logged, or included in any error value or debug representation
[N-14].

Implementations expose exactly six storage operations---lookup by selector,
insert, mark-rotated, revoke-if-active, revoke-family, revoke-user [N-16]---so
that production adapters over relational or key-value stores are mechanical.
Two of the six are compare-and-set operations rather than unconditional
writes, for reasons developed in \S\ref{sec:cas}.

\paragraph{Retention is normative.} A record must be retained, with its
status and successor pointer intact, until at least its family expiry
[N-15]. This was operational advice in an earlier draft of the
specification and is now a requirement, because it is load-bearing: reuse
detection \emph{is} the act of finding a \textit{rotated} record. A store
that expires rotated rows early---a Redis TTL applied at rotation, a nightly
sweep of non-active rows, an ORM's cascade on session cleanup---silently
converts every replay from \texttt{REUSE\_DETECTED} into \texttt{NOT\_FOUND},
disabling the single property the design exists to provide, with no error, no
log line, and no failing test. Records may be deleted once the family
deadline has passed.

\subsection{Refresh Algorithm}
\label{sec:refresh}

Token exchange performs the ten checks of Table~\ref{tab:checks} in fixed
order [N-26]. Step~4 compares
$\mathrm{HMAC}(\mathit{pepper}_{record.kid}, \mathit{verifier})$ to the
stored hash in constant time with an explicit length guard [N-31]; step~9
evaluates sender binding with the \emph{record's} pepper, so that pepper
rotation cannot invalidate bound sessions [N-32]; step~10 mints the
successor under the \emph{active} pepper, migrating the device hash forward
[N-33].

Fixing the order fixes the observable error surface, which both simplifies
clients and prevents divergent information leakage across implementations.
The ordering is not merely a convenience: it is the reason a \textit{rotated}
record presented with a \emph{wrong} verifier returns
\texttt{VERIFIER\_MISMATCH} and revokes nothing [N-28]. Were reuse handling
placed before the verifier proof, knowledge of a selector---a value the
design deliberately treats as public---would suffice to destroy an arbitrary
family through the reuse path.

\begin{table}[t]
\centering\small
\caption{The ten refresh checks, in normative order [N-26]. The order is
observable: it fixes which code is returned when several conditions hold at
once, and is itself pinned by the five behavioural scenarios whose identifiers
begin \texttt{order-01} to \texttt{order-05}. Steps marked \ddag\ perform a
compare-and-set whose
failure is itself a protocol outcome (\S\ref{sec:cas}). $^\dagger$At the
default $g = 0$ step~5 always yields \texttt{REUSE\_DETECTED}; where a grace
window is enabled the retry it admits can instead end in
\texttt{DEVICE\_MISMATCH} or \texttt{CONFLICT} (\S\ref{sec:reuse}). A deployment is
expected to collapse all of these to one generic response at the transport
boundary and log the specific code server-side [N-42].}
\label{tab:checks}
\begin{tabular}{@{}c l l l@{}}
\toprule
\# & Check & On failure & Observable code \\
\midrule
1 & Parse the presented string & reject before any store call & \texttt{MALFORMED} \\
2 & Pepper exists for token \texttt{kid} & reject & \texttt{UNKNOWN\_KID} \\
3 & Record exists for selector & reject & \texttt{NOT\_FOUND} \\
4 & Constant-time verifier proof & reject; family untouched & \texttt{VERIFIER\_MISMATCH} \\
5 & Status is not \textit{rotated} & revoke family, or grace retry & \texttt{REUSE\_DETECTED}$^\dagger$ \\
6 & Status is not \textit{revoked} & reject & \texttt{REVOKED} \\
7 & Family deadline not passed & revoke family & \texttt{EXPIRED\_ABSOLUTE} \\
8 & Idle deadline not passed & revoke family & \texttt{EXPIRED\_IDLE} \\
9 & Sender binding matches & revoke family & \texttt{DEVICE\_MISMATCH} \\
10\,\ddag & Rotate under compare-and-set & revoke inserted successor & \texttt{CONFLICT} \\
\bottomrule
\end{tabular}
\end{table}

\subsection{Concurrent Rotation and the Compare-and-Set Contract}
\label{sec:cas}

The transition of the presented record from \textit{active} to
\textit{rotated} is the one place in the design where reuse detection can be
bypassed with no cryptographic failure whatsoever, and it is the change we
consider most consequential since the earliest drafts of this specification.

Suppose that transition is an unconditional write. Two concurrent refreshes
of the same active token (T5) both execute step~3, both read status
\textit{active}, both pass step~4, and both proceed to step~10. Each mints a
successor, each inserts it, and each marks the predecessor rotated. Both
callers receive a valid token. The family has forked into two independently
valid lineages, each rooted at a record that believes itself the sole
successor, and---this is the damaging part---no subsequent presentation of
either lineage is a replay. Reuse detection is not weakened; it is switched
off for that family, permanently and silently. The condition is reachable by
an ordinary retrying client, and it was reproducible against an earlier,
unpublished draft of this specification, in which the transition was an
unconditional write.

The store contract closes it by making the transition a compare-and-set
[N-17]. The \texttt{markRotated} operation takes the expected
\texttt{fromStatus} as an argument, applies its write if and only if the
stored record's current status equals it, and returns whether the write
applied. In SQL this is an
\texttt{UPDATE \ldots{} WHERE selector = ? AND status = ?} with the
affected-row count returned; a store that returns \texttt{true}
unconditionally is non-conforming. Exactly one of two racing refreshes can
therefore win. The loser has already inserted a successor, so it revokes
that successor and returns \texttt{CONFLICT} [N-34]: a transient outcome
that revokes nothing else, that never accompanies a token, and that clients
retry once, the retry then meeting the ordinary reuse path [N-35].

\begin{figure}[t]
\centering
\begin{tikzpicture}[
  >=Stealth,
  st/.style={draw, rounded corners=2pt, minimum width=21mm, minimum height=8mm,
             font=\small\ttfamily},
  cas/.style={very thick},
  lb/.style={font=\scriptsize, align=center, inner sep=2.5pt}]

  \node[st] (a) at (0,0) {active};
  \node[st] (r) at (6.4,0) {rotated};
  \node[st] (v) at (3.2,-3.0) {revoked};

  \draw[<-] (a.west) -- ++(-1.0,0) node[lb, left] {\texttt{issue}};

  \draw[->, cas] (a) -- (r)
    node[lb, midway, above=1mm] {\texttt{refresh}: insert successor,\\
      then \texttt{markRotated(active)}\,\ddag};

  \path (r) edge[->, cas, loop above, min distance=8mm, out=115, in=65] (r);
  \node[lb, above=13mm of r] {grace retry within $g$:\\
    \texttt{revokeIfActive(succ)}\,\ddag\ then\\
    \texttt{markRotated(rotated)}\,\ddag};

  \draw[->] (a) -- (v)
    node[lb, midway, left=3mm] {\texttt{revokeIfActive}\,\ddag\\
      \texttt{revokeFamily}\\ \texttt{revokeUser}\\ expiry, device mismatch};

  \draw[->] (r) -- (v)
    node[lb, midway, right=1mm] {replay outside $g$:\\
      \texttt{REUSE\_DETECTED},\\ whole family revoked\\
      (also \texttt{revokeFamily} / \texttt{revokeUser})};
\end{tikzpicture}
\caption{Record lifecycle. The operations marked \ddag\ are the two
compare-and-set writes: they apply only if the record's current status matches
the expected one, and they report whether they applied. The remaining
transitions into \textit{revoked} are unconditional bulk writes. A lost
compare-and-set yields
\texttt{CONFLICT} and mints nothing (\S\ref{sec:cas}). \textit{revoked} is
terminal, and every state is retained until the family's absolute deadline
[N-15]---deleting \textit{rotated} records early is what disables reuse
detection. A grace retry re-rotates in place and preserves the original
rotation time, so the window $g$ is anchored to the first rotation and cannot
be extended.}
\label{fig:states}
\end{figure}
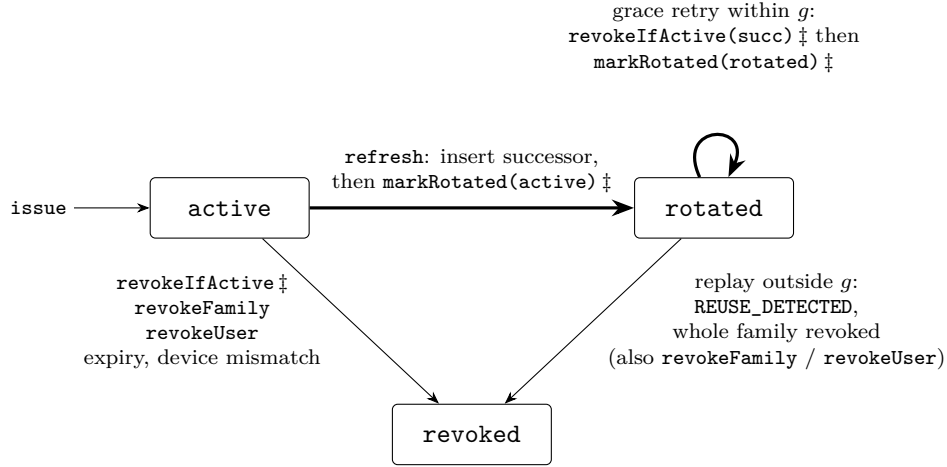

The same mechanism protects the grace path at a second point: revocation of
the unused successor during a lost-response retry is itself a compare-and-set
(\texttt{revokeIfActive}, [N-18]), so two simultaneous retries cannot both
re-rotate; the loser returns \texttt{CONFLICT} without minting anything
[N-30]. Figure~\ref{fig:states} marks both compare-and-set edges.

Two obligations remain outside the engine. First, the store should execute
the rotation write pair---insert of the successor, then compare-and-set of
the predecessor---inside one transaction [N-22]; where it cannot, the
engine's compensation is what prevents a lost race from leaving a live
orphan record. Second, the atomicity of the compare-and-set is delegated to
the store, and is only as strong as the store provides.

\subsection{Replay Handling and Bounded Retry}
\label{sec:reuse}

Presentation of a \textit{rotated} record is, by default, conclusive evidence
that two parties hold one credential; the entire family is revoked and a
\texttt{REUSE\_DETECTED} event is surfaced for monitoring. One legitimate
counterexample exists: a client that issued a refresh and never received the
response. NEBULA admits an optional \emph{grace window} of configurable
duration $g$, disabled by default: a rotated record presented within $g$ of
its \emph{original} rotation time, whose successor exists and is still
unused, and whose family has not passed its absolute deadline, is treated as
a lost-response retry---the unused successor is revoked and the chain
re-rotated [N-30]. Three properties are normative: the window is anchored to
the first rotation time and is therefore \emph{not extendable} by repeated
retries; any evidence of successor use converts the retry into a theft
verdict; and a retry can never mint a token past the family's absolute
deadline.

\paragraph{The window costs detectability, not just margin.} This is the part
that is easy to state too gently, so we state it as the specification now
does. While $g > 0$, an adversary holding a rotated predecessor who acts
within $g$ of the rotation and \emph{before} the legitimate client uses its
successor satisfies every grace-retry condition. The unused successor is
revoked, the adversary is served a live token, and the legitimate client's
next refresh returns \texttt{REVOKED}. \textbf{No \texttt{REUSE\_DETECTED}
event is raised.} Note which interleaving this is: with $g = 0$ it is
precisely the case reuse detection catches most cleanly---the victim
refreshed first, and the thief replayed a rotated token. Enabling the window
converts that case from a detected compromise into a silent takeover
observable only as an unexplained \texttt{REVOKED}.

The specification therefore requires deployments to treat
$g$ as a reliability-versus-detectability trade rather than a tuning
parameter, recommends the smallest value clients need, recommends leaving
the default of zero unless lost-response retries are an observed problem,
and recommends alerting on the \texttt{REVOKED} rate wherever the window is
enabled [N-30]. This reconciles, in specified form and with its cost stated,
the tension between strict replay detection and unreliable networks that
providers currently resolve ad hoc.

\subsection{Authenticated Revocation}
\label{sec:revoke}

Revocation by token is an \emph{authenticated} operation [N-36]:
\texttt{revokeToken} performs steps~1--4 of Table~\ref{tab:checks}---parse,
pepper lookup, record lookup, constant-time verifier proof---and returns the
corresponding failure when any of them fails. Only then does it revoke the
family. It succeeds regardless of the record's status, so that a client can
log out with a token that has already been rotated or revoked.

The requirement exists because the selector is a \emph{public lookup key by
construction}. It is the only token-derived value the specification permits
to be indexed [N-45], it is explicitly permitted as a log correlation
identifier [N-46], and it is exactly what a database dump yields in the
clear. A value with those properties must not also be a capability. An
earlier revision of this design accepted a selector alone for revocation,
which made every one of those disclosure channels an unauthenticated way to
terminate any session---denial of service against an arbitrary user, from
material the rest of the design goes to some length to render inert.
Administrative revocation, which the application authenticates itself, is
served instead by the family- and user-scoped operations, which take a
server-side identifier and no token [N-37].

\subsection{Key Rotation}

Peppers are named by kid; verification always uses the pepper named by the
\emph{record}, while minting always uses the configured \emph{active}
pepper. Introducing a new pepper and switching the active kid therefore
rotates keys with zero invalidation: outstanding tokens verify under their
recorded kid and migrate to the new pepper at their next natural rotation.
A record written under a pepper that has since been retired yields
\texttt{UNKNOWN\_KID} rather than a spurious proof failure [N-27].

\section{Security Analysis}
\label{sec:security}

\paragraph{Forgery (T3).} Two cases. Absent any corresponding database
record, acceptance requires guessing both a stored selector and the 256-bit
verifier that belongs to it. Given a known selector---which the design
treats as public---acceptance requires supplying a verifier whose
HMAC-SHA-256 tag under the record's pepper equals the stored one. That is
\emph{existential forgery against a MAC under an unknown key}, not second
preimage on an unkeyed hash: HMAC is keyed, the adversary does not hold the
pepper, and the tag is therefore not a function the adversary can even
evaluate, let alone invert. HMAC is unforgeable under standard assumptions
on its compression function~\cite{bellare2006,rfc2104,fips198}, and peppers
are held outside the database by construction, so the reduction is to pepper
secrecy rather than to any property of SHA-256 alone.

\paragraph{Token theft (T1).} A stolen token is a bearer credential until its
next use, which NEBULA cannot prevent and does not claim to. The analysis
concerns the aftermath, and it turns on which party refreshes first.

If the victim refreshes first, the stolen token is rotated; the attacker's
presentation of it is a replay, the family is revoked, and the attacker never
authenticates---except within an enabled grace window, where this is exactly
the silent-takeover case of \S\ref{sec:reuse}.

If the attacker refreshes first, they obtain the successor and the victim
retains the rotated predecessor. The victim's next refresh is then the
replay, and it revokes the family, terminating the attacker's lineage. The
attacker's window is bounded by the victim's time to next refresh, which in
an \emph{active} session is at most one access-token lifetime.

It is not bounded that way in an abandoned session, and the distinction
matters. Idle expiry does not contain an attacker who is using the session,
because [N-33] recomputes
$\mathit{idleExpiresAt} = \min(\mathit{now} + \mathit{idleTtl},
\mathit{familyExpiresAt})$ on \emph{every} successor: the attacker's own
refreshes slide the idle clock forward indefinitely, and a victim who never
returns never triggers the replay. For a session the victim abandons, the
only effective bound is the family's absolute deadline, which is fixed at
login and never extended by anything. That deadline is the sole non-slidable
limit in the scheme, and it is the parameter a deployment must choose with
this case in mind; relying on idle expiry to bound a compromise is a
mistake. The comparison with non-rotating deployments is still favourable
(there, the bound is the full remaining token lifetime with no detection at
any point), but the improvement in the abandoned-session case is detection
and lineage containment, not a shorter window.

Optional device binding rejects the attacker's \emph{first} use when the
binding input is unavailable to them, with strength proportional to how
strongly that input is bound to the device: strong for keystore-held
identifiers on native platforms, weaker on the web, where cryptographic
sender-constraining should be preferred.

\paragraph{Concurrent rotation (T5).} Without the compare-and-set contract of
\S\ref{sec:cas}, two concurrent refreshes of one token fork the family into
two valid lineages and permanently disable reuse detection for it, with no
cryptographic failure and no observable error. With it, exactly one refresh
wins; the loser revokes the successor it inserted and reports
\texttt{CONFLICT}, which revokes nothing else and is safe to retry. We
highlight this because it is the failure mode a policy-level document is
least likely to surface: it is invisible to any single-threaded test, it
produces no error at the time it occurs, and its consequence---the absence
of a future detection---is not observable at all until an attacker relies on
it. Two behavioural scenarios---\texttt{conflict-01-lost-rotation-cas} and
\texttt{conflict-02-lost-grace-cas}---drive both losing paths by forcing the
compare-and-set to report failure, so
an implementation that ignores the returned flag---minting a token, or
leaving the orphan successor active---cannot pass conformance. What those
scenarios pin is the engine's half of the contract: that it consults the
result and compensates. The store's half, that \texttt{markRotated} applies
the \texttt{fromStatus} predicate rather than writing unconditionally, is
not observable through a runner-supplied store and remains an obligation on
the adopter's implementation.

\paragraph{Database disclosure (T2).} Records contain only HMAC outputs under
peppers held outside the database. An adversary with the full table cannot
reconstruct a single presentable token, and device identifiers are likewise
not recoverable, which additionally serves data-minimization obligations.

Non-actionability has a second component that is easy to lose, and that this
design lost once: it must also hold for \emph{revocation}. Because the
selector is a public lookup key by construction (\S\ref{sec:revoke}), a
revocation interface accepting a selector alone would convert every table
row, and every log line, into an unauthenticated ability to terminate that
session. [N-36] closes this by requiring the verifier proof for revocation
by token; [N-28] closes the parallel path through reuse handling by ordering
the verifier proof before it. Together they ensure that a disclosed database
yields neither authentication nor denial of service.

\paragraph{Timing (T4).} Database lookups key only on the public selector
[N-45]; secret-derived comparisons are constant-time with length guards and
are specified never to throw on malformed input [N-8, N-31], removing
exception-timing oracles.

\paragraph{Post-quantum posture.} NEBULA contains no public-key cryptography,
so Shor's algorithm~\cite{shor1997} has no target anywhere in the design---in
contrast to asymmetric-signed token formats, whose signing keys it would
recover. The applicable quantum attack is Grover search~\cite{grover1996}.
Against the 256-bit verifier it costs approximately $2^{128}$ queries,
conventionally regarded as post-quantum safe, and that figure is
unconditional here because the verifier is 32 bytes of CSPRNG output by
construction.

The corresponding statement for the HMAC layer is \emph{not} unconditional
and we do not assert it as such. Recovering a pepper by search costs
$2^{n/2}$ for a pepper of $n$ bits of entropy, not $2^{128}$ by default, so
the margin holds only where the pepper carries close to 256 bits. That is
precisely why [N-23] recommends that peppers be generated by a CSPRNG with
at least 256 bits of entropy and held in an environment variable,
secret manager, or KMS, and why the specification describes its minimum
pepper length as a floor against obvious misconfiguration rather than a
sufficient condition. A deployment using a 32-character human-chosen
passphrase satisfies the length floor and does not satisfy this claim.

Because validity is server-side state, harvest-now-decrypt-later strategies
are inapplicable in a stronger sense than for any ciphertext: a recorded
token is inert once rotated, expired, or revoked, independently of future
computational advances. These statements concern the token layer only;
surrounding transport and any co-deployed asymmetric tokens carry their own
migration obligations.

\paragraph{Non-goals.} A compromised server holding peppers and store access
defeats any token scheme, as does persistent client compromise; NEBULA
documents both explicitly rather than claiming their mitigation. Rate
limiting is likewise out of scope and is required of the deployment at the
edge.

\section{Implementations and Cross-Language Conformance}
\label{sec:conformance}

The specification is implemented in ten languages---TypeScript, Python, Go,
Rust, Java, PHP, C\#, Ruby, Elixir, and Dart---each as a single small package
with an injectable clock, the six-method storage contract, and an in-memory
reference store documented as unsuitable for production [N-21]. The Java
package is plain JDK bytecode and is callable from any JVM language; we test
it from Java only, and make no claim of tested Kotlin or Scala support.

\paragraph{Dependencies, counted rather than asserted.} Eight of the ten
packages carry no runtime dependency at all beyond the platform library,
taking their cryptographic primitives from
\texttt{node:crypto} (TypeScript); \texttt{hmac}, \texttt{hashlib} and
\texttt{secrets} (Python); \texttt{crypto/hmac}, \texttt{crypto/rand},
\texttt{crypto/sha256} and \texttt{crypto/subtle} (Go);
\texttt{javax.crypto} with \texttt{MessageDigest} and
\texttt{SecureRandom} (Java); \texttt{hash\_hmac} and
\texttt{hash\_equals} (PHP);
\texttt{System.Security.Cryptography} (C\#); \texttt{openssl} and
\texttt{securerandom} (Ruby); and Erlang's \texttt{:crypto} (Elixir). Two do
not: Rust takes six crates (\texttt{hmac}, \texttt{sha2}, \texttt{base64},
\texttt{hex}, \texttt{rand}, \texttt{subtle}) and Dart takes
\texttt{package:crypto}, in both cases because the language ships no keyed
hash in its standard library---Dart does ship a CSPRNG, \texttt{Random.secure},
which is what the port uses. Test-only and build-time
dependencies---JSON parsers for the shared vector files, test frameworks, and
a source-link generator in the C\# build---are excluded from that count, as
none reaches a consumer.

\paragraph{Conformance artefacts.} Conformance is defined operationally, by
published artefacts rather than by prose (Table~\ref{tab:artifacts}).
Deterministic \emph{test vectors} fix the keyed primitives and the parser:
expected HMAC outputs for verifier and device hashing under published
peppers, and 32 accept/reject cases for token parsing covering oversize
input, wrong arity, empty parts, bytes outside the alphabet (including
padding and the standard-base64 characters \texttt{+} and \texttt{/}), wrong
part lengths, and all three non-canonical encodings of a known verifier.
These pin byte-level agreement across languages.

\paragraph{The behavioural suite is data, not prose.} This is the part we
would emphasise as method. \texttt{spec/behavior-vectors.json} carries 38
scenarios; each is an identifier, the list of requirements it exercises, and
a sequence of operations against a fixed clock with expected outcomes and
expected resulting record states. The file also fixes the small operation
vocabulary a runner must implement: issue, refresh, revoke by token, revoke
by family, revoke by user, advance the clock, rebuild the engine over the
same store with different peppers, force a named compare-and-set to fail
once, assert the multiset of record statuses in the store, and assert that no
stored field contains any issued token, verifier, or raw device identifier.
Each implementation supplies a thin runner over that vocabulary and executes
the shared file; no implementation restates the scenarios in its own test
framework.

The alternative---one suite ported case-for-case into each language---is
what this project did first, and it failed in the predictable way: the ports
drifted in case count before anyone noticed, because each language's suite
was green against its own copy and nothing compared the copies. Publishing
the suite as a single artefact removes that failure mode rather than
policing it, and it makes third-party conformance mechanical: a port
demonstrates conformance by running the file, not by re-deriving it [N-49].
The counts are themselves normative---a runner must assert that the number
of cases it executed equals the published count, and must fail if a section
is absent or empty [N-48]---so silently iterating zero cases is a
conformance failure rather than a pass. Thirty-seven of the 38 scenarios are
unconditional; the single exception is guarded by a named runtime condition
(whether the language's string type can represent an unpaired surrogate),
and a runner must report by identifier any conditional scenario it skips.

\begin{table}[t]
\centering\small
\caption{The published conformance artefacts. Every implementation in the
repository---and any third-party port claiming conformance---executes the
two vector files directly, through a per-language runner that adapts the
shared operation vocabulary to the local API.}
\label{tab:artifacts}
\begin{tabular}{@{}l l r@{}}
\toprule
Artefact & What it pins & Cases \\
\midrule
\texttt{SPECIFICATION.md}
  & Normative requirements, individually citable [N-\emph{n}] & 53 \\[2pt]
\texttt{spec/test-vectors.json}
  & Verifier hashing: HMAC output, byte for byte & 7 \\
  & Device hashing: HMAC output over \texttt{"device:"}\,$\|$\,$d$ & 7 \\
  & Token parsing: accept/reject boundary of the grammar & 32 \\[2pt]
\texttt{spec/behavior-vectors.json}
  & Engine behaviour against an injected clock & 38 \\
  & \quad of which unconditional & 37 \\[2pt]
\texttt{spec/traceability.json}
  & Requirement $\rightarrow$ vector/scenario mapping & 53 \\
  & \quad cited by a vector or scenario & 23 \\
  & \quad verified by review or by runner assertion & 30 \\
\bottomrule
\end{tabular}
\end{table}

\paragraph{What this establishes, and what it does not.} It establishes that
the ten implementations agree \emph{at the observation points the vectors
define}: the byte-level output of both keyed hashes, the accept/reject
boundary of the parser, and---for each scenario, over whichever of them that
scenario asserts---the returned outcome code, the returned generation, the
family and expiry relations, and the resulting multiset of record statuses in
the store. Agreement is enforced by those
shared artefacts rather than by intention or by review. The published
traceability map (Table~\ref{tab:artifacts}) covers all 53 requirements: 23
are cited by a vector or a scenario, and the remaining 30---API shape,
extensibility policy, documentation, logging and deployment obligations---are
verified by review, which we publish as an honest label rather than suppress
as a gap. Three of that thirty are a boundary case we name rather than round
off: the constants block [N-4] is checked against the vectors' own
\texttt{constants} section by every per-language conformance runner; clock
injection [N-3] is what all 38 behavioural scenarios are driven through; and
each of the ten codes of the error set [N-38] is the expected outcome of at
least one scenario. All three are therefore executed without being cited by a
vector.

It does not establish behavioural indistinguishability in general. The
vectors do not observe timing, memory behaviour, log output, or store call
ordering beyond what individual scenarios assert, and no differential
harness yet drives all ten implementations from one randomly generated
operation transcript and compares their complete outputs. We name that
harness as future work rather than claim its result.

We report this less as engineering detail than as method: for security
mechanisms whose failure modes live in edge-case semantics, publishing the
conformance suite as an executable artefact alongside the prose
specification collapses the gap---endemic to informal standards---between
what the document says and what implementations do.

\section{Limitations and Future Work}
\label{sec:limitations}

NEBULA's guarantees are argued, not machine-verified. Formal modelling of the
rotation state machine and of its compare-and-set transitions---in
Tamarin~\cite{tamarin} or ProVerif~\cite{proverif}---an independent security
audit, and fuzzing of the ten parsers would, we believe, be prerequisites to
any strong deployment claim. None of the three is scheduled; they are named so
that a reader can weigh what is missing, not as a roadmap. Cross-language agreement is
currently established only at the observation points the shared vectors
define (\S\ref{sec:conformance}); the differential harness described there
would extend it over a far larger input space and does not yet exist. The
design accepts first use of a stolen token absent device binding; a
cryptographic sender-constraining profile binding families to DPoP-style
proof-of-possession keys is the natural extension. The implementations,
while conformant under the shared suite, have no production deployment
history at the time of writing. Finally, the storage contract
presumes a store with read-after-write consistency on the rotation path, and
the compare-and-set of \S\ref{sec:cas} is only as strong as the store's own
atomicity; characterizing behaviour under weaker consistency models remains
open.

\section{Conclusion}

The refresh-token layer occupies an unusual position: universally
implemented, security-critical, mandated in outline by current best practice,
and specified in detail by no one. NEBULA fills that gap with a small
normative specification of 53 individually citable requirements, an argument
for its security properties---including a concurrency contract without which
reuse detection can be switched off silently, and an incidental but
consequential post-quantum robustness---and ten implementations whose
agreement is enforced by executable conformance artefacts rather than by
intention. The specification, the implementations, and the conformance
artefacts are published under the Apache License 2.0. We offer the result as
both a deployable mechanism and a template for specifying the other quietly
divergent layers of practical authentication.

\end{document}